# Second-Harmonic Generation and Spectrum Modulation by Active Nonlinear Metamaterial

Zhiyu Wang[1,2], Yu Luo[1,2], Liang Peng[1,2], Jiangtao Huangfu[1,2], Tao Jiang[1,2],

Dongxing Wang[3], Hongsheng Chen[1,2], Lixin Ran[1,2]

[1]*The Electromagnetic Academy at Zhejiang University, Zhejiang University, Hangzhou 310027*

[2]*Department of Information and Electronic Engineering, Zhejiang University, Hangzhou 310027*

[3]*School of Engineering and Applied Sciences, Harvad University, Cambridge, MA 02138*

## Abstract

The nonlinear properties of a metamaterial sample composed of double-layer metallic patterns and voltage controllable diodes are experimentally investigated. Second harmonics and spectrum modulations are clearly observed in a wide band of microwave frequencies, showing that this kind of metamaterial is not only tunable by low DC bias voltage, but also behaves strong nonlinear property under a small power incidence. These properties are difficult to be found in normal, naturally occurring materials.

In 1999, J. Pendry *et. al.* predicted that the split ring resonators (SRRs) could lead to a magnetism at radio frequency (RF) or even optical frequency range [1]. This property is a prerequisite for the design of the negative-index metamaterials, which attracted substantial attentions in the communities of physics and electromagnetics later on [2-5]. Another predication made by Pendry's group is that they showed theoretically that the enhanced nonlinear electromagnetic (EM) properties could arise from such metamaterials, and this was confirmed by M. Klein *et. al.* in 2006 [6]. Afterwards, the properties of nonlinear metamaterials have been widely studied [7-12]. Different kinds of tunable metamaterials have been proposed. Most of the designs are based on the mechanism that introduce some kind of controllable material or lumped elements to commonly used metamaterial unit cells, like SRRs, to construct an active metamaterial. In the optical frequency band, a photosensitive semiconductor is used [13]. In the microwave frequency band, some ferroelectric/ferromagnetic composite metamaterial (CMM) are realized, such as $Ba_{0.5}Sr_{0.5}TiO_3$ (BST, ferroelectric) [14] and $Y_3Fe_5O_{12}$ (YIG, Yttrium Iron Garnets, ferromagnetic) [15-17]. These metamaterials can be controlled by lights, high voltages or dynamic magnetic fields, correspondingly. In order to make a new active metamaterial which is more convenient to control, the nonlinear lumped element, varactor diode, which can be tuned just by low direct current (DC) voltages, is also considered. Several researches have been conducted by mounting diodes in metamaterial resonators [18-21]. In this letter, we introduce a new resonant pattern of such kind of active metamaterial which has both electric and magnetic nonlinear effects. As well as it could be tuned by low DC voltages, it demonstrates strong nonlinear effects. Under the illumination of a low power incident EM wave, the phenomena of second-harmonic generation, signal demodulation and intermodulation are clearly observed from only one piece of the board made of such metamaterial. This indicates the great potential to apply nonlinear metamaterial in information processing area.

We incorporate a microwave voltage controllable diode into a double-layer metallic resonant pattern, shown in top-left of figure 1, as the unit cell to construct an active metamaterial. The pattern is I-shaped and all patterns are printed in alignment on both sides of a dielectric subtract. In each unit cell, two varactor diodes are soldered in the middle gap of the pattern on each layer with the bias voltage added on the four strips along $y$ axis. For an incident electric field polarized along the x axis, the electric resonance is induced by the conductive currents flowing in the strips along the x axis on both layers and the magnetic resonance can also be induced by the resonant current flowing along the same strips and the two capacitance structures sandwiched by the strips along the y axis, when the $y$ component of magnetic field exists in the substrate. Therefore, such structure supports simultaneous electric and magnetic resonance, and, thanks to the existence of the diodes, will behave both electric and magnetic nonlinear properties.

Figure 1 shows the concrete realization and the dimensions of the basic unit cell that will be investigated in this letter. In the realization of the experimental sample, we repeat such patterns along two orthogonal directions to obtain a single layer slab with a spatial periodicity of 6 mm along the *x* and *y* directions, respectively. The cupreous patterns are printed on an 1-mm-thick FR4 substrate with a relative permittivity around 4.6. The number of unit cells on one board is $48 \times 40$. Infineon BAT15-03W Diodes with a small surface mount package are selected to serve as the controllable nonlinear lumped element. Dimensions of one unit cell is shown in the bottom-right of figure 1, where l = h = 6 mm, g = 1.6 mm, $w_1$ = 0.3 mm and $w_2$= 1 mm. The magnified image of one part of the sample soldered with diodes is shown in the bottom-left of figure 1.

We use a DC source to control the bias voltage of the diodes, choosing the strongly nonlinear region on the volt-ampere characteristic curve of the diodes (see in the datasheet of Infineon BAT15-03W) to obtain the nonlinear behavior with a low power incidence. In order to reduce the maximum DC voltage needed, we change the orientation of the diodes by a period of 4 unit cells when soldering them, such that the whole board is divided into 12 identical zones, and each zone only needs one forth of the original DC voltages to control the diodes.

The experiments for measuring the nonlinear behaviors of the metamaterial sample are organized in three steps. The first step is to verify the voltage controllable property of the sample. The experimental setup is shown in the inset of figure 2. An EM wave with a power of 20 dBm from a vector signal generator (Agilent E8267C) is fed into a wide-band horn antenna to illuminate the metamaterial sample, and the transmitted power is measured by a same type antenna connecting with a spectrum analyzer ( Agilent 4407B) on the opposite side. In the experiment, we scan the bias voltage for different incident frequencies and record the corresponding transmission coefficient (S21) in figure 2. Since the effective capacitances of the varactor diodes change with the bias voltages, we can find that the transmission power through the single layer sample changes sharply with the bias voltage below the frequency 10.15 GHz, which is nearly the upper limit of the working frequency of Infineon BAT15-03W. At frequencies higher than 10.15 GHz, the choosing diode will act as a fixed effective capacitance, losing its original function of varactor, therefore the transmission power will not have distinct change any more (Data not shown). So, figure 2 shows that the transmission property of the sample is well controlled by the bias voltages.

In step 2, we try to measure the harmonic components to verify the existence of the nonlinearity of the sample using the same experimental setup and input power as in step 1. Since Infineon BAT15-03W is designed for X-band applications, we tune the frequency of the input monolithic wave between 8 to 12 GHz, and the results for 4 different bias voltages at 9 GHz are recorded in figure 3, where the second harmonics are clearly observed. In the experiment, the amplitude of the second-harmonic component increases with the increase of the bias voltage until the bias voltage applied to each four diodes is 0.9 V, which is the point where strongest nonlinear is observed, and then it starts to decrease. The magnified zone of second-harmonic components is shown in the middle inset of figure 3. The frequencies of higher order harmonics have exceeded the working frequency of the antennas, therefore their amplitudes are small and are not shown here.

In step 3, trying to observe the spectrum modifications caused by the nonlinearity of the sample, two different monolithic inputs denoted by $f_1$ and $f_2$ are fed simultaneously into the source antenna, shown in figure 4(a), and the components at demodulation frequency, i.e., $f_2 - f_1$, and the intermodulation frequencies, i.e., $2f_1 - f_2$ and $2f_2 - f_1$, are expected to appear. The corresponding results are shown in figure 4(b) and (c), respectively. In figure 4(b), a low frequency component about 800 MHz is clearly observed, which is the exact difference of 9.2 GHz ($f_1$) and 10 GHz ($f_2$), indicating that a distributed demodulation has occurred while the EM wave passing through the sample. In figure 4(c), the intermodulation components at 9 GHz and 10.5 GHz when monolithic inputs are 9.5 GHz ($f_1$) and 10 GHz ($f_2$), respectively, are also clearly seen, which precisely correspond to the 3rd-order intermodulation frequencies calculated by $2f_1 - f_2$ and $2f_2 - f_1$, respectively.

Similarly, the amplitudes of the modulation and intermodulation components change as the same manner as in step 1 when bias voltage is changed, showing strongest nonlinearity at 0.9 V.

The aforementioned experiments clearly verified that the metamaterial sample is not only tunable by low DC bias voltage, but also behaves strong nonlinear property under small power incidence. These properties are difficult to be found in normal, naturally occurring materials, therefore should have plenty of potential applications, such as in distributed signal processing and receiving. For the frequency range we have used in this letter, i.e., from 3.65 GHz to 10.15 GHz in figure 2, the corresponding wavelengths are from about 82 mm to 29.6 cm, and the periodicity of the unit cells in the metamaterial sample is 6 mm, which is in most cases less than one fifth or more of the input wavelengths, meaning that such an artificial sample can be regarded as an effective media. In linear case, the constitutive relation is $\overline{D} = \varepsilon \overline{E}$, $\overline{B} = \mu \overline{H}$, therefore we use effective permittivity and permeability to describe the EM properties of a linear metamaterial. However, in nonlinear case, the relation turns to $\overline{D} = \varepsilon \overline{E} + \chi_e^{(1)} \overline{E}^2 + \cdots$, $\overline{B} = \mu \overline{H} + \chi_m^{(1)} \overline{H}^2 + \cdots$, therefore we may also use $\chi_e$ and $\chi_m$ to describe the nonlinearity of nonlinear metamaterial. We will report this in a separate paper.

**References**

[1] J. B. Pendry, A. J. Holden, D. J. Robbins, and W. J. Stewart, "Magnetism from conductors and enhanced nonlinear phenomena", IEEE Trans. Microw. Theory Tech. **47**, 2075 (1999).

[2] R. A. Shelby, D. R. Smith, S. Schultz, "Experimental Verification of a Negative Index of Refraction", Science **292**, 77 (2001).

[3] T. J. Yen, W. J. Padilla, N. Fang, D. C. Vier, D. R. Smith, J. B. Pendry, D. N. Basov, and X. Zhang, "Terahertz Magnetic Response from Artificial Materials", Science **303**, 1494 (2004).

[4] S. Linden, C. Enkrich, M. Wegener, J. Zhou, T. Koschny, and C. M. Soukoulis, "Magnetic Response of Metamaterials at 100 Terahertz", Science **306**, 1351 (2004).

[5] A. N. Grigorenko, A. K. Geim, H. F. Gleeson, Y. Zhang, A. A. Firsov, I. Y. Khrushchev, and J. Petrovic, "Nanofabricated media with negative permeability at visible frequencies", Nature **438**, 335 (2005).

[6] M. W. Klein, C. Enkrich, M. Wegener, and S. Linden, "Second-Harmonic Generation from Magnetic Metamaterials", Science **313**, 502 (2006).

[7] A. A. Zharov, I. V. Shadrivov, and Yu. S. Kivshar, "Nonlinear properties of left-handed metamaterials," Phys. Rev. Lett. **91,** 037401–4 (2003).

[8] M. Lapine, M. Gorkunov, and K. H. Ringhofer, "Nonlinearity of a metamaterial arising from diode insertions into resonant conductive elements," Phys. Rev. E **67,** 065601–4 (2003).

[9] M. Gorkunov and M. Lapine, "Tuning of a nonlinear metamaterial band gap by an external magnetic field," Phys. Rev. B **70,** 235109–9 (2004).

[10] S. Lim, C. Caloz, and T. Itoh, "Metamaterial-based electronically controlled transmission-line structure as a novel leaky-wave antenna with tunable radiation angle

and beamwidth," IEEE Trans. Microwave Theory Tech. **52,** 2678–2690 (2004).

[11] H. T. Chen, W. J. Padilla, J. M. O. Zide, A. C. Gossard, A. J. Taylor, and R. D. Averitt, "Active terahertz metamaterial devices," Nature (London) **444,** 597–600 (2006).

[12] I. V. Shadrivov, S. K. Morrison, and Yu. S. Kivshar, "Tunable split-ring resonators for nonlinear negative-index metamaterials," Opt. Express **14,** 9344–9349 (2006).

[13] A. Degiron, J. J. Mock, and D. R. Smith, "Modulating and tuning the response of metamaterials at the unit cell level," Opt. Express **15,** 1115–1127 (2007).

[14] L. Peng, L. Ran, H. Chen, H. Zhang, J. A. Kong, and T. M. Grzegorczyk, "Experimental observation of left-handed behavior in an array of standard dielectric resonators," Phys. Rev. Lett. **98**, 157403 (2007).

[15] Y. He, P. He, S. D. Yoon, P.V. Parimi, F.J. Rachford, V.G. Harris, C. Vittoria, "Tunable negative index metamaterial using yttrium iron garnet," Journal of Magnetism and Magnetic Materials **313**, 187–191 (2007).

[16] H. Zhao, L. Kang, J. Zhou, Q. Zhao, L. Li, L. Peng, and Y. Bai, "Experimental demonstration of tunable negative phase velocity and negative refraction in a ferromagnetic/ferroelectric composite metamaterial," Appl. Phys. Lett. **93**, 201106 (2008).

[17] L. Kang, Q. Zhao, H. Zhao, and J. Zhou, "Magnetically tunable negative permeability metamaterial composed by split ring resonators and ferrite rods," Opt. Express **16**, 8825-8834 (2008).

[18] D. A. Powell, I. V. Shadrivov, Y. S. Kivshar, and M. V. Gorkunov, "Self-tuning mechanisms of nonlinear split-ring resonators," Appl. Phys. Lett. **91,** 144107 (2007).

[19] D. Wang, L. Ran, H. Chen, and M. Mu, J. A. Kong and B.-I. Wu, "Active

left-handed material collaborated with microwave varactors," Appl. Phys. Lett. **91**, 164101 (2007).

[20] I. V. Shadrivov ,A. B. Kozyrev, and D. W. van der Weide, Yu. S. Kivshar "Tunable transmission and harmonic generation in nonlinear metamaterials," Appl. Phys. Lett. **93**, 161903 (2008).

[21] I. V. Shadrivov ,A. B. Kozyrev, and D. W. van der Weide, Yu. S. Kivshar "Nonlinear magnetic metamaterials," Opt. Express **16**, 20266-20271 (2008).

**Figure captions**

Fig. 1. Photograph of the metamaterial sample used in this letter. Geometry of the double-layer metamaterial pattern is on the top-left, with its dimensions on the bottom-right.

Fig. 2. Transmission coefficient as functions of the bias voltage for different frequencies.

Fig. 3. Output spectrum with second-harmonic components. Input monolithic wave is at 9 GHz. Second-harmonic components at 18 GHz is shown at different bias voltages.

Fig. 4. (a) Setup of the spectrum modulation experiment. (b) Output spectrum of demodulation signal. The difference frequency component around 800 MHz is shown with two input monolithic waves at 9.2 GHz and 10 GHz, respectively. (c) Output spectrum with 3rd-order intermodulation components. Two intermodulation components exist at 9 GHz and 10.5 GHz, with two input monolithic waves at 9.5 GHz and 10 GHz, respectively.

Figure 1

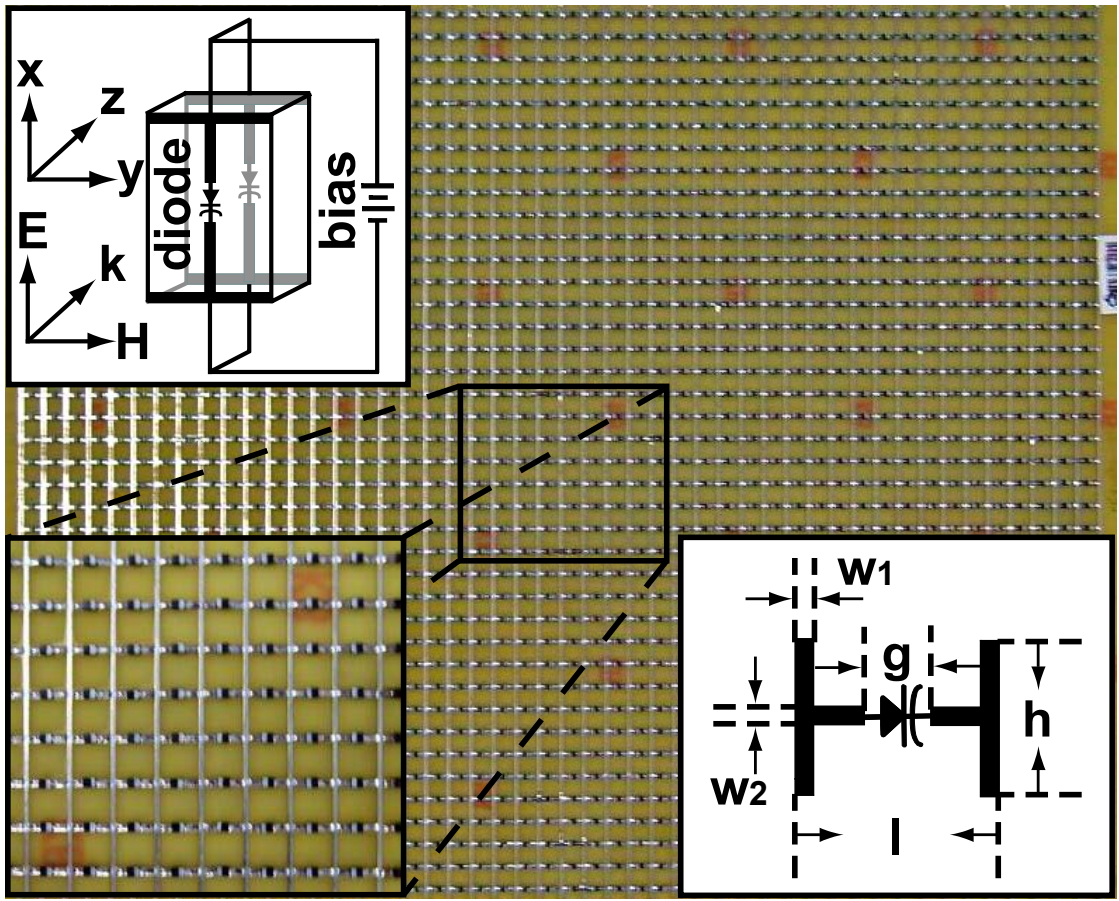

Figure 2

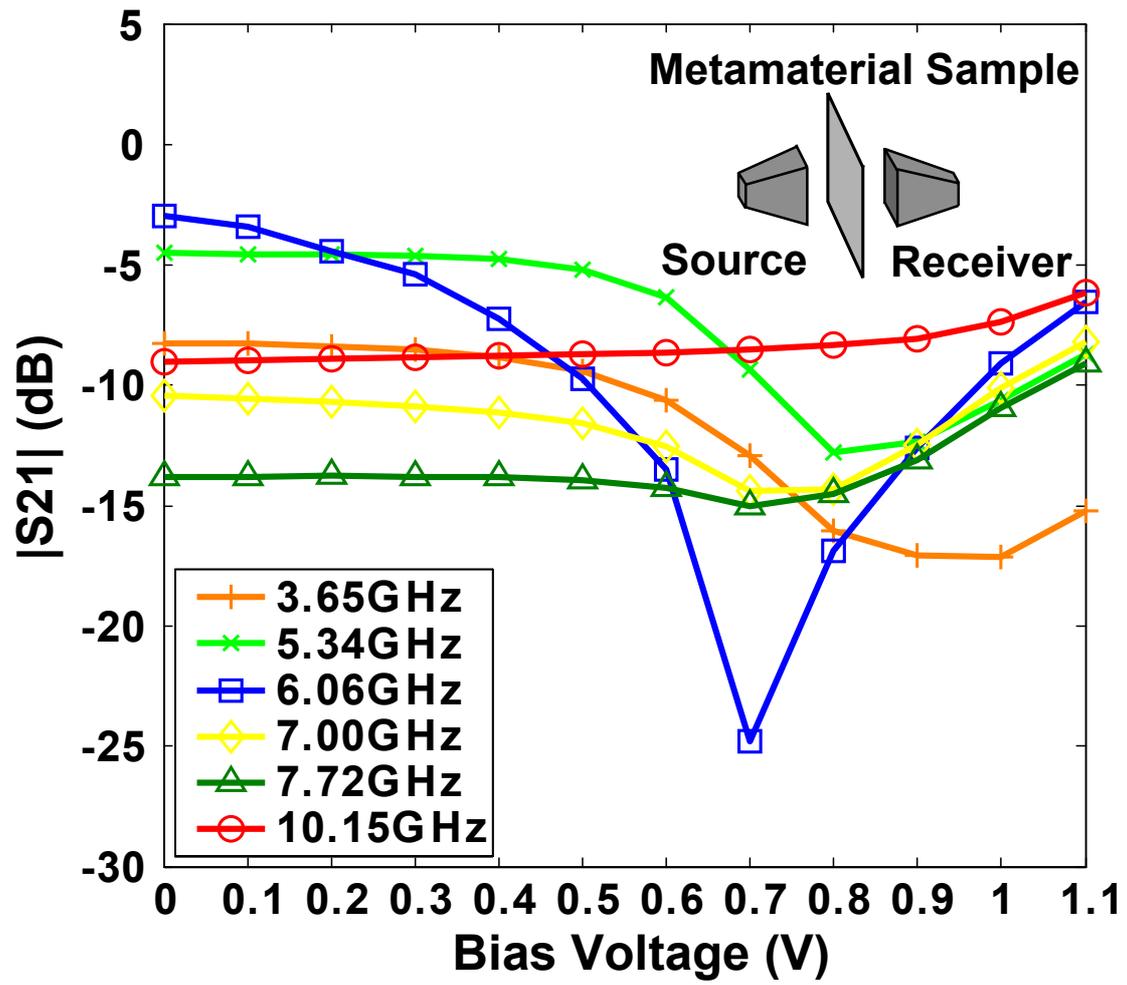

Figure 3

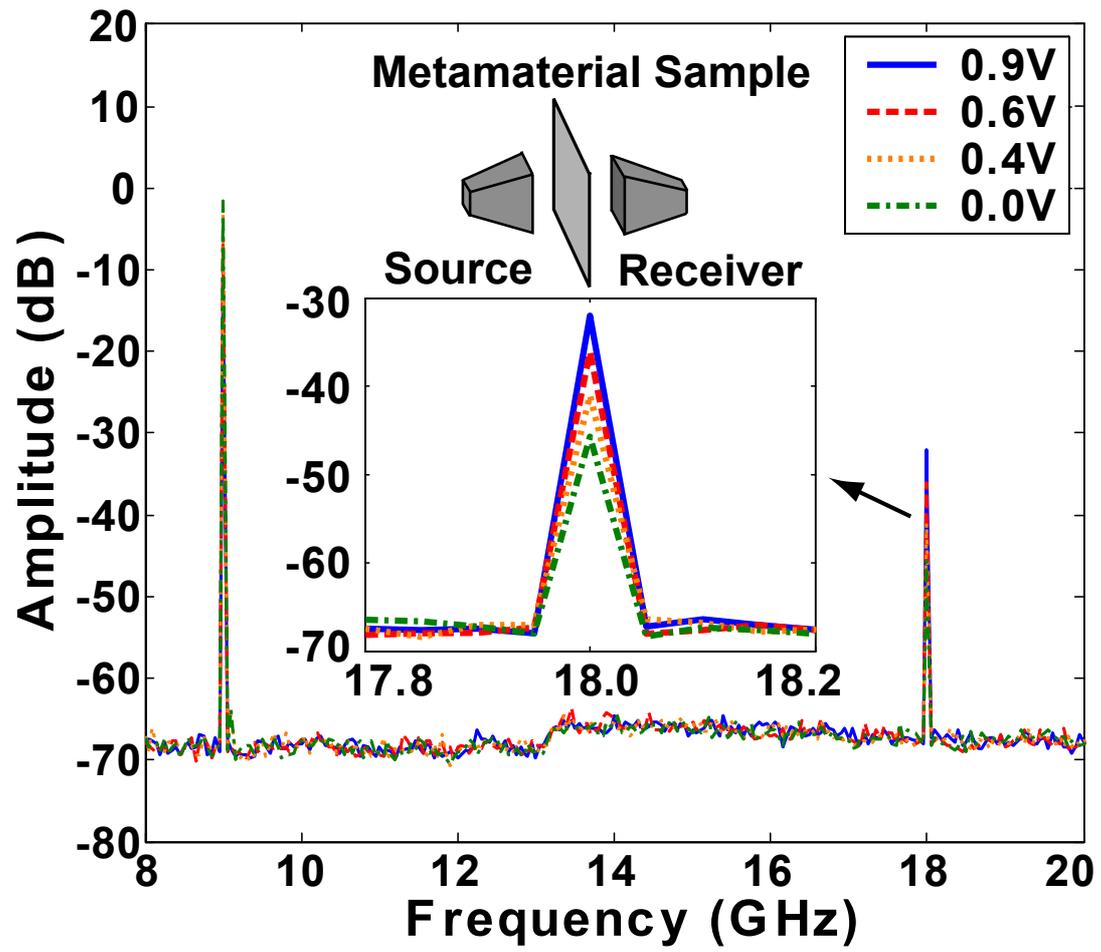

Figure 4

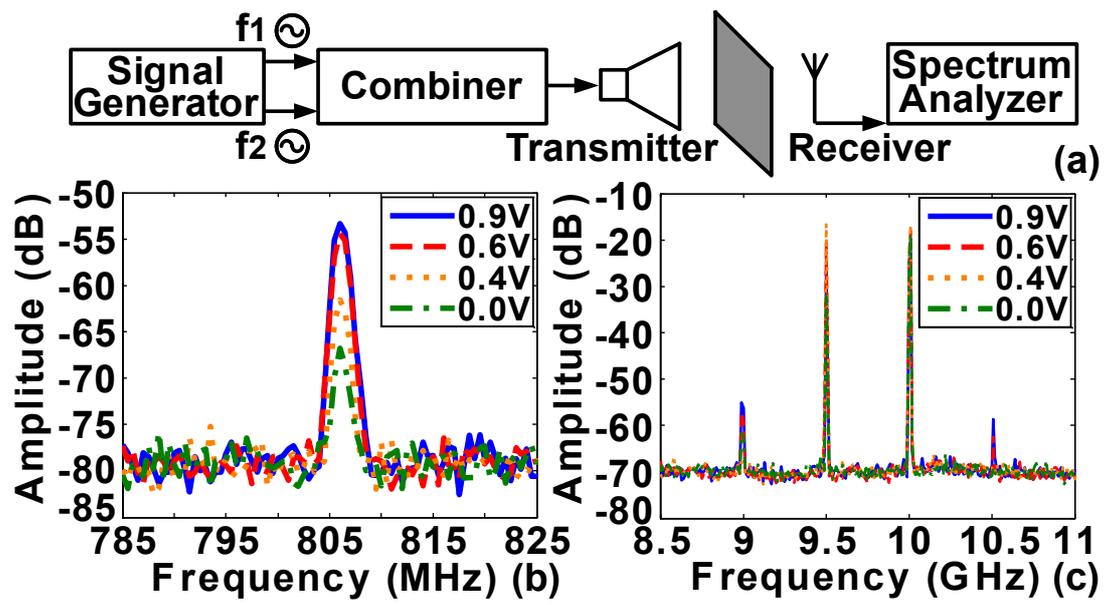